# Model for processive movement of myosin V and myosin VI

Ping Xie, Shuo-Xing Dou, and Peng-Ye Wang

*Laboratory of Soft Matter Physics, Institute of Physics, Chinese Academy of Sciences,*
*P .O. Box, 603, Beijing 100080, China*

## Abstract

Myosin V and myosin VI are two classes of two-headed molecular motors of the myosin superfamily that move processively along helical actin filaments in opposite directions. Here we present a hand-over-hand model for their processive movements. In the model, the moving direction of a dimeric molecular motor is automatically determined by the relative orientation between its two heads at free state and its head's binding orientation on track filament. This determines that myosin V moves toward the barbed end and myosin VI moves toward the pointed end of actin. During the moving period in one step, one head remains bound to actin for myosin V whereas two heads are detached for myosin VI: The moving manner is determined by the length of neck domain. This naturally explains the similar dynamic behaviors but opposite moving directions of myosin VI and mutant myosin V (the neck of which is truncated to only one-sixth of the native length). Because of different moving manners, myosin VI and mutant myosin V exhibit significantly broader step-size distribution than native myosin V. However, all three motors give the same mean step size of ~ 36 nm (the pseudo-repeat of actin helix). Using the model we study the dynamics of myosin V quantitatively, with theoretical results in agreement with previous experimental ones.

Among the superfamily of myosin, class V and class VI myosins have been demonstrated to be able to move processively along helical actin filaments [1–7]. Myosin V moves toward the barbed ends while myosin VI toward the pointed ends of actin filaments [8, 9]. Processivity means that one molecule can undergo multiple ATPase cycles and coupled mechanical steps for each diffusional encounter with actin. This is adapted obviously for their cellular functions in organelle and vesicular transport. The opposite moving directions imply their distinct roles. Both myosin V and myosin VI are composed of two heavy chains, each consisting of a motor domain (MD), a neck domain, an *α*-helical coiled coil and a globular target-binding domain [8–10]. One of the main structural difference of myosin V and myosin VI is in their neck domains: The neck of myosin V has six light-chain binding sites (IQ motifs) bound with six calmodulins, while that of myosin VI has only one IQ motif bound with one calmodulin. Consequently the neck of myosin VI is much shorter than that of myosin V. Another structural difference is that myosin VI has two small insertions of 9 and 13 residues in a region between the nucleotide pocket and the actin binding interface of the MD and has one large insertion of 53 residues between the IQ motif and the converter domain in the MD.

Now there is no consensus as to the mechanism of the processive movement of myosin V and myosin VI. The presently prevailing model for myosin V is the hand-over-hand model which is based on the lever-arm mechanism [2, 5, 11–15]. In that model, it is supposed that the motor maintains continuous attachment to actin by alternately repeating single-headed and double-headed binding. The model requires that the two heads move in a *coordinated* manner and alternately move past each other. The model can explain the experimental results on the dynamics such as average step size (~36 nm). Especially, a recent elegant experiment by Yildiz *et al.* [15], in which they used high-resolution single fluorophore imaging technique to track a single fluorophore attached to one of the long myosin heads as it moves along actin, provides compelling evidence in favor of the hand-over-hand model. However, it is difficult to explain with the model the experimental observations on mutant myosin V by Tanaka *et al.* [16]: Although the neck of myosin V is truncated to only one-sixth of the native length, surprisingly they found that the mutant myosin V can still make processive movements along actin and, in addition, the average step size is still ~36 nm! In addition, in order to explain the large step size of myosin VI [6, 7] with the model, it has to be assumed that myosin VI operates by a combination of a short power-stroke of the short lever arm (neck domain) coupled with a significantly extended conformation adopted by the 53-residue large insertion [6]. However, as just mentioned, mutant myosin V with only one IQ but no such an insertion can still make large steps!

An alternative model based on biased Brownian motion has been proposed for myosin V and myosin VI [3, 7]. There the movement of a myosin is realized by biased thermal diffusion of its heads along actin. Thus the neck length does not affect the step sizes. The model is supported by the experimental results with myosin VI and mutant myosin V. However, it is not consistent with the experimental observation for myosin V by Yildiz *et al.* [15]. In addition, the origin of the potential slope is not clear.

For quantitatively studying the dynamics of molecular motors, two approaches are often used. One is the multistate chemical kinetic description [17, 18]. In this approach, it is postulated that a motor protein molecule steps through a sequence of discrete chemical states linked by rate constants. The other one is the thermal ratchet model in which a molecular motor is viewed as a Brownian particle moving in two (or more) periodic but spatially asymmetric stochastically switched potentials [19–22].

In this paper, we propose a new hand-over-hand model that can explain not only the processive movement of native myosin V, but also that of myosin VI and mutant myosin V. According to the model, the moving direction of a dimeric motor is determined by the relative orientation between the two heads at free state of the motor and its head's binding orientation on track filament; while its mean step size is determined by the pseudo-repeat of actin helix. The model does not require the two heads' coordination in their ATPase and mechanical cycles. Rather, the ATPase activities at the two heads are independent. During the moving period in one step, one head is bound to actin for myosin V, whereas both heads are detached for myosin VI. Using the model, the rather puzzling experimental results of mutant myosin V [16] become readily understandable: Mutant myosin V has a very similar dynamic behavior but opposite moving direction compared with myosin VI. The significantly greater spread of step-size distribution of myosin VI (or mutant myosin V) than that of native myosin V is easily explained. The theoretical results of the dwell duration between steps versus load for different ATP concentrations and the dwell duration versus [ATP] at low load for myosin V are in agreement with the previous experimental results [1].

**1. Myosin V**

We assume that the free state of myosin V is as shown in Fig. 1(a) (left) and its MD binds to the binding site on actin in an orientation as shown in Fig. 1(a) (right). Since the coiled coil that connects the two neck domains contains no IQ motif and calmodulin, the distance between the two

MD's can vary easily within a limited range around their equilibrium distance. This is consistent with the electron microscopy observations [23].

Because each head of myosin V has a long neck domain, containing six IQ motifs plus six calmodulins, it is reasonable to assume that the neck is elastically bendable. Thus the two heads can bind simultaneously to two equivalent motor binding sites (~36 nm apart) along actin in equivalent orientations [Fig. 2(a)]. Note that the neck shapes in Fig. 2(a), which are determined by the internal elastic force and torque resulted from the change of the orientation of the right MD from that in Fig. 1(a), are consistent with the observations by electron microscopy [24]. Once one of the two heads or both become(s) free, the internal elastic force and torque drives the dimeric myosin V to its free (or equilibrium) state [Fig. 2(b)]. This is based on the principle of minimum free energy, for the equilibrium state has the minimum free energy and the rigor state [Fig. 2(a)] has a greater free energy.

The structural studies of myosin heads and actin helix filaments by X-ray crystallography reveals that a myosin MD, for example, of a vertebrate smooth muscle myosin has +16 net elementary charges [25] and a monomer of actin has − 5 net elementary charges [26]. Thus we assume that the binding force between a myosin MD and the actin filament is electrostatic. When the MD is very close to a binding site of actin, the electrostatic binding force is dependent on the charge distributions on the two surfaces [27, 28], thus resulting in fixed binding orientation of the MD on actin [Fig. 1(a) (right)].

The conformational change of a myosin head at its actin-binding site results in the variation of the electrostatic binding force between the myosin head and actin. For the weakening of the binding force we make the following hypothesis. When there is a conformational change caused by ATP binding or ATP hydrolysis, there appear some gaps between the MD and actin. Therefore, the solvent will flow into the gaps, causing the relative dielectric constant in the gaps, $\varepsilon_r^{(gap)} = 1 + \chi_e$, to increase from the low ε of solute [27]. As the MD which attracts negative ions around it departs from its binding site on actin, there will be more positive ions than negative ions left in the local solvent near the binding site, which results in the increase of screening effect on the negative charges of the nearby actin monomers. This is equivalent to that the value of $\varepsilon_r^{(local)}$ of the local solvent can be much larger than the average value of $\varepsilon_r$ of the solvent for a certain time, and then $\varepsilon_r^{(local)}$ relaxes to $\varepsilon_r$.

Here we give an illustration of whether ATP binding or ATP hydrolysis results in the weakening of the binding force. Using Tryp fluorescence De La Cruz *et al.* [29] determined an ATP hydrolysis rate of $\geq 750$ s$^{-1}$ for myosin V, and Trybus *et al.* [10] experimentally determined a myosin dissociation rate of $\sim 850$ s$^{-1}$ by ATP binding. Both rates are nearly the same and are much higher than the ATPase cycle rate of $12-15$ s$^{-1}$ [29]. Therefore, either hypothesis that ATP binding promotes the MD detachment or that ATP hydrolysis leads to the weakening of binding force is consistent with the experimental result. In the case of kinesin, which performs the same task of intracellular transport and is in the same solvent environment as myosin V, it is the ATP hydrolysis that results in the weakening of binding force [30].

The proposed mechanism for the processive movement of myosin V is similar to that of kinesin [31]. We begin with two heads bound to the two binding sites (*i.e.*, rigor state) by the electrostatic binding forces, as shown in Fig. 2(a) or (a'). According to the moving direction, we will call the head close to the barbed end the leading head and that close to the pointed end the trailing head for myosin V.

We consider two cases for ATP binding:

(i) ATP binds to the trailing head earlier. The increase of $\varepsilon_r^{(local)}$ of the local solvent induces the electrostatic binding force to become smaller than the internal elastic force. Thus the trailing head is detached and then driven to its equilibrium position [Fig. 2(b)], the trailing head becoming the new leading head (ADP-$P_i$ bound). Then the electrostatic force from some binding site of actin near the new leading head induces the head coming close to the binding site. $P_i$ is then rapidly released upon contacting of the detached head with actin, and then the strong binding force overcomes the elasticity of retaining the equilibrium state, driving the MD to its binding orientation [Fig. 2(c)]. Thus a forward step is made with one ATP being hydrolyzed (1:1 coupling).

(ii) ATP binds to the leading head earlier. The leading head is then detached and the internal elastic force and torque drive the leading head to its equilibrium state [Fig. 2(b')]. After $\varepsilon_r^{(local)}$ of the local solvent relaxes to $\varepsilon_r$ of the solvent, the detached head is bound [as in case (i)] to the binding site again by the recovered binding force [Fig. 2(c')]. Then ATP binds to the trailing head and is hydrolyzed, and myosin V makes a forward step as in case (i). Two ATP molecules are consumed to make this forward step (2:1 coupling). After this step ATP has a larger probability to bind to the new trailing head earlier than the new leading head if the ATPase cycle rates at the two heads are assumed to be the same.

In fact, in the rigor state as shown in Fig. 2(a) there exist a forward internal elastic force on the trailing head and a backward force on the leading head. As will be discussed in detail in Sec. 4, this results in that the ATPase rate at the trailing head is much higher than the leading head. Therefore, myosin V generally hydrolyzes one ATP per step (1:1 coupling).

**2. Myosin VI**

We assume that the free state of myosin VI is as shown in Fig. 1(b) (left) and its MD's binding orientation is as shown in Fig. 1(b) (right). Note the different relative orientation between two heads of myosin VI [Fig. 1(b) (left)] from that of myosin V [Fig. 1(a) (left)] and mutant myosin V [Fig. 1(c) (left)]. The difference may be due to the presence of small insertions in the MDs of myosin VI. This conjecture is based on the experiment by Homma *et al*. [9], where by constructing various chimaeric myosins with different combinations of the MDs and necks from myosin V and myosin VI they found that it is the MD that determines the direction of myosin movement.

Because myosin VI has a short neck domain, containing only one IQ motif plus one calmodulin, it will be much more rigid to bending than the neck of myosin V. In another word, it requires much stronger force and torque to change the relative orientation between the two heads from that as shown in Fig. 1(b) (left) to that the two heads are in equivalent binding orientations. Therefore, once a head is bound to a binding site of actin, we assume that the other head cannot bind to some nearby binding site and still remains at its equilibrium position. This is different from myosin V as mentioned in Sec. 1 and from kinesin [31]. In the case of kinesin, although the neck linker of a head is also short, because there is no calmodulin binding, it is considered to be elastically stretchable/compressible and bendable. In addition, the binding force of microtubule is stronger than that of actin, and this stronger binding force can overcome the internal force and torque and drive the unbound kinesin head bound to microtubule. That is, at rigor state, two kinesin heads are bound to microtubule.

The proposed mechanism for the processive movement of myosin VI is described as follows. We begin with one head (in nucleotide-free state) of myosin VI bound to a binding site of the actin filament. As required by the equilibrium state the unbound head (in ADP-P$_i$–bound state) will be in the position as shown in Fig. 3(a). According to the moving direction, we will call the head close to the pointed end the leading head and that close to the barbed end the trailing head for myosin VI (noting the difference from myosin V).

ATP binding at the bound myosin head leads to the increase of $\varepsilon_r^{(local)}$ of the local solvent, resulting in the weakening of the electrostatic binding force of the nearby actin-binding sites. Now the electrostatic force between the positively charged leading head and the negatively-charged binding sites is asymmetric in the horizontal direction, as shown in Fig. 3(b), where the binding site with the strongest electrostatic binding force is shown in red and as the color becomes shallower the electrostatic binding force becomes weaker. Therefore, the horizontal component $F_E$ of the net electrostatic force acting on the leading head points to the pointed end of actin. On the other hand, because the present electrostatic force between the trailing head and the actin-binding sites is symmetric, the horizontal component of the net electrostatic force acting on the trailing head is zero. Thus driven by $F_E$ from the leading head, the trailing head is detached from the weak binding site [Fig. 3(b)], and then myosin VI diffuses toward the pointed end [Fig. 3(c)]. Note that there always exists the vertical component of the net electrostatic force, acting on myosin VI, that points toward actin during diffusion and this prevents myosin VI from diffusing away from actin.

Since now both heads are detached, even a weak binding force can make the leading head bind to some nearby binding site of actin in the fixed orientation with P$_i$ release [Fig. 3(d)]. The binding probability follows the bound level distribution (which depends on the electrostatic binding force) as measured by Mehta *et al.* [32] and Steffen *et al.* [33] and, therefore, the step size of myosin VI has a broad distribution similar to that of the bound level. Once the leading head binds to actin, the trailing head will immediately become the new leading head as required by the equilibrium state [Fig. 3(d)]. The new bound head is now in ADP–bound state. Activated by actin, ADP is released and the trailing head becomes nucleotide-free [Fig. 3(e)]. One mechanochemical cycle is completed, with one ATP being hydrolyzed per step (1:1 coupling).

It is noted that the moving behaviour of myosin VI is different from that of myosin V. (i) For myosin V there is always one head bound to actin and the two heads can bind to actin simultaneously; whereas for myosin VI both heads can be detached from the actin simultaneously but can never bind to actin simultaneously. The moving behavior of the former is like human walking; whereas that of the latter is somewhat like human running. (ii) Myosin V moves processively to the barbed end of actin; whereas myosin VI moves processively to the pointed end. The moving directions of dimeric motors, including kinesin, are determined by the relative orientations between the two heads at free states of the motors and their heads' binding orientations on the track filaments.

It is interesting to note that, if the neck domain of myosin V is mutantly truncated to only one-sixth of the native length, its moving behaviour and dynamics should become the same as those of myosin VI except the moving direction. The relative orientation between the two heads of mutant myosin V at free state [Fig. 1(c) (left)] and its head's binding orientation on actin [Fig. 1(c) (right)], which are the same as those for native myosin V, determine its processive movement to the barbed end. This is consistent with the experimental results of Tanaka *et al.* [16]: The step-size distribution is the same as that of myosin VI [6, 7], but its moving direction is opposite to myosin VI.

## 3. Moving time in one step

In this section we will calculate the moving time $\tau$ of myosin V and myosin VI in one step.

To this end, we resort to the following equation for an over-damped Brownian particle

$$\Gamma\, dx/dt = F_{driving} - F_{load} + f(t)\,,\tag{1}$$

where $\Gamma$ is the frictional drag coefficient, $v = dx/dt$ the moving velocity of the particle along actin, $F_{driving}$ the driving force, and $F_{load}$ the force by load. $f(t)$ is the fluctuating Langevin force, with $\langle f(t)\rangle = 0$ and $\langle f(t)f(t')\rangle = 2k_B T \Gamma\, \delta(t-t')$.

**Myosin V.** The moving time $\tau$ is defined as the time for the detached head moving from binding site (I) to binding site (III) in Fig. 2. From the Stokes formula we have $\Gamma = 6\pi\eta r_k = 9.4\times 10^{-11}\,\mathrm{kg\,s^{-1}}$, where the viscosity $\eta$ of the aqueous medium of a cell around myosin V is approximately $0.01\,\mathrm{g\,cm^{-1}\,s^{-1}}$ and the myosin head is approximated as a sphere with radius $r_k \approx 5\,\mathrm{nm}$. The driving force is assumed to be equal to the stall force, $F_{driving} = 3\,\mathrm{pN}$ [1]. From Eq. (1) we obtain that the mean first-passage time $T$, i.e., the moving time $\tau$, for the detached head to travel a distance of $L = 2d$ has the form (see Appendix A)

$$T = \frac{1}{F_d}\left[L + \frac{D}{F_d}\left(\exp(-\frac{F_d}{D}L) - 1\right)\right]\,,\tag{2}$$

where $F_d = (F_{driving} - F_{load})/\Gamma$, $D = k_B T/\Gamma$, and $d$ is the actin helix pseudo-repeat (~ 36 nm). From Eq. (2) we obtain the moving time $\tau$ versus load as shown in Fig. 4. It is seen that $\tau \approx 2\,\mu\mathrm{s}$ at $F_{load} = 0$. With the increase of $F_{load}$ the moving time $\tau$ increases. Since the experimental values of $\tau$ are not available now, we cannot make a direct comparison between the calculated and measured values. However, the calculated values shown in Fig. 4 are in the same range as the measured ones of $0 \sim 50\,\mu\mathrm{s}$ for kinesin [34]. At stall force, i.e., $F_{load} = F_{driving}$, $\tau$ becomes infinity.

**Myosin VI.** The moving time $\tau$ is defined as the time for myosin VI to diffuse from the position as shown in Fig. 3(b) to that in Fig. 3(d) driven by force $F_{driving} = F_E$. Because there are two heads we take $\Gamma = 12\pi\eta r_k = 1.88\times 10^{-10}\,\mathrm{kg\,s^{-1}}$. The driving force is assumed to be equal to the stall force, $F_{driving} = 2.8\,\mathrm{pN}$ [6]. Using Eq. (2) and letting $L = d$ we obtain the moving time $\tau$ versus load as shown in Fig. 4. It is seen that the values of $\tau$ for myosin VI are close to those for myosin V.

## 4. Dynamics

As we noted in the above section, the moving times of myosin V and myosin VI in one step are several orders shorter than ATPase cycle time (~100 ms [29]). Thus, during one ATPase cycle both heads are almost always bound to actin for myosin V and one head bound to actin for myosin VI. That means that the moving velocities of myosin V and myosin VI actually only depend on the ATPase rates at two heads in rigor state in Fig. 2(a) for myosin V and in Fig. 3(a) for myosin VI, respectively.

To calculate ATPase rate $K$ at a head, we define the mean ATP binding rate, $k_b = k_{b0}[\mathrm{ATP}]$, and the mean ATP turnover rate, $k_c$. Then $K$ can be written as (see Appendix B)

$$K = \frac{k_c[\mathrm{ATP}]}{[\mathrm{ATP}] + k_c/k_{b0}}\,.\tag{3}$$

For a more precise calculation of *K*, we should consider the probability distributions of ATP binding time and ATP turnover time instead of their mean values $t_b = 1/k_b$ and $t_c = 1/k_c$. For this purpose, we define $g(t)$ as representing the probability distribution of the time interval between two successive collisions of ATP molecules with the ATP binding site of a myosin head and $h(t)$ as representing the probability distribution of ATP turnover time, *i.e.*, the time interval from ATP binding to ADP release. Usually, $g(t)$ and $h(t)$ are independent and they take the following forms

$$g(t) = \frac{a_1 a_2}{a_2 - a_1}\left[\exp(-a_1 t) - \exp(-a_2 t)\right], \quad (a_2 > a_1 > 0) \tag{4a}$$

$$h(t) = \frac{k_1 k_2}{k_2 - k_1}\left[\exp(-k_1 t) - \exp(-k_2 t)\right]. \quad (k_2 > k_1 > 0) \tag{4b}$$

Thus the mean ATP binding time is $t_b \equiv \frac{1}{k_b} = \frac{a_1 + a_2}{a_1 a_2}$, and the mean ATP turnover time $t_c \equiv \frac{1}{k_c} = \frac{k_1 + k_2}{k_1 k_2}$. The two-exponential distributions [Eqs. (4a) and (4b)] are consistent with the measured distributions of dwell time for myosin V (Fig. 3 of Ref. [2]). Note that at high ATP concentrations the dwell-time distribution is approximately determined by $h(t)$; whereas at low ATP concentrations it is approximately determined by $g(t)$. From the experiment, we know that $k_2 \gg k_1$ and $a_2 \gg a_1$.

From Eqs. (4a) and (4b) we obtain the mean ATPase cycle time $\overline{T}$ (see Appendix B)

$$\overline{T} = \frac{a_1 + a_2}{a_1 a_2} + \frac{a_1 a_2}{a_2 - a_1}\frac{1}{k_2 - k_1}\left\{\frac{k_2}{k_1^2}\left(\varsigma\left[2,\frac{k_1 + a_1}{k_1}\right] - \varsigma\left[2,\frac{k_1 + a_2}{k_1}\right]\right) - \frac{k_1}{k_2^2}\left(\varsigma\left[2,\frac{k_2 + a_1}{k_2}\right] - \varsigma\left[2,\frac{k_2 + a_2}{k_2}\right]\right)\right\},$$
(5)

where $a_1 = a_1^{(0)}[\text{ATP}]$, $a_2 = a_2^{(0)}[\text{ATP}]$. The mean ATPase rate is

$$K = 1/\overline{T}. \tag{6}$$

When there is a force *F* exerted on the MD in the moving direction of myosin V, we have (see Appendix B)

$$k_1 = \frac{k_{10}(1 + A_c)}{1 + A_c \exp(-F\delta/k_B T)}, \tag{7a}$$

$$k_2 = \frac{k_{20}(1 + A_c)}{1 + A_c \exp(-F\delta/k_B T)}, \tag{7b}$$

$$a_1^{(0)} = \frac{a_{10}(1 + A_b)}{1 + A_b \exp(-F\delta/k_B T)}, \tag{7c}$$

$$a_2^{(0)} = \frac{a_{20}(1 + A_b)}{1 + A_b \exp(-F\delta/k_B T)}. \tag{7d}$$

**Myosin V.** When there is a load $F_{load}$ in the opposite direction of myosin V movement, the forces exerted on MD's of the two heads in rigor state can be calculated by referring to Fig. 5, where the neck length and the angles between necks and actin are taken in accordance with experimental results [35]. Thus the forces on the two MD's in the motor's movement direction can be written as

$$F_{trailing} = (F_0 - F_{load})\cos 35°, \qquad (8a)$$

$$F_{leading} = -F_0 \cos 35° - F_{load} \cos 52°, \qquad (8b)$$

where $F_0$ is the internal elastic force. It should be noted that we use the net internal force $F_0$ on the leading head in Eq. (8b) for simplicity.

Since the ATPase rate at the trailing MD is much faster than the leading MD (which will be seen below), the dwell duration between steps, $t_d$, (or the moving velocity $V$) is essentially only dependent on the ATPase rate $K_{trailing}$ at the trailing MD, i.e., $t_d = 1/K_{trailing}$. Using Eqs. (5)-(8a), the experimentally measured dwell duration $t_d$ versus load (Fig. 2d in Ref. [1]) can be fitted well, which is shown in Fig. 6(a). It is noted that from the fitted parameter values $k_{10} = 0.25$ s$^{-1}$ and $k_{20} = 3$ s$^{-1}$ we have $k_1 = 10.3$ s$^{-1}$ and $k_2 = 123$ s$^{-1}$ at low loads (where load and $t_d$ are uncorrelated), which are very close to the experimental values of $k_1 = 12.5$ s$^{-1}$ and $k_2 = 150$ s$^{-1}$ for dwell-time distribution at saturating [ATP] = 2 mM (Fig. 3*A* of Ref. [2]). Similarly, from $a_{10} = 0.72$ μM$^{-1}$s$^{-1}$ and $a_{20} = 3.6$ μM$^{-1}$s$^{-1}$ we have $a_1 = 1.6$ s$^{-1}$ and $a_2 = 8$ s$^{-1}$ at low [ATP] = 2 μM and low loads, which are close to the experimental values of $a_1 = 2.7$ s$^{-1}$ and $a_2 = 13.8$ s$^{-1}$ for dwell-time distribution at the same [ATP] (Fig. 3*C* of Ref. [2]). For comparison, in the figure we also show the result (dashed curve) calculated by using Eqs. (3), (7), and (8a). Note that at [ATP] = 2 mM the dashed curve coincides with the solid curve. We see that the theoretical results by using probability distributions of ATP binding time and ATP turnover time are in better agreement with the experimental results than those by using their mean values. Using the parameter values in Fig. 6(a) we calculate $t_d$ versus [ATP], which is shown in Fig. 6(b). The theoretical result is in agreement with the experimental one at low loads [1]. In Fig. 7 we show the calculated ATPase rates versus [ATP] at the two heads for $F_{load} = 0$. It can be seen that the ATPase rate at the trailing head is much faster than the leading head, which means that myosin V generally hydrolyzes one ATP per step (1:1 coupling).

Because the distance between the two MD's of myosin V can vary within a limited range around their equilibrium distance, and, on the other hand, the binding force of the binding sites on actin must be strong enough to overcome the elasticity of retaining the equilibrium orientation of the detached head, it is anticipated that the detached head will bind to actin in a narrow range. Thus the step-size distribution is also narrow. However, because myosin VI (or the mutant myosin V in Ref. [16]) diffuses with both heads detached, it is anticipated that, similar to the bound-level distributions as observed in Refs. [32] and [33], the leading head binds to actin in a wide range. Thus the step-size distribution is significantly broader than that of native myosin V. Because of the diffusion behavior, we also expect that myosin VI (or the mutant myosin V) may have backward steps against forward force $F_E$ and have forward steps of two-step size. These deductions are in agreement with the experimental observations [6, 7, 16].

At last, because during the moving period one head remains bound to actin for myosin V whereas two heads are detached from actin for myosin VI, we would expect that it is relatively much easier for myosin VI to diffuse away from (or release) actin than myosin V. That means that myosin V can take much more steps than myosin VI before releasing actin. This is supported by experiments: Myosin V can take at least 40 - 50 steps on average before releasing actin [1], whereas we see that, from the measured mean run length of 226 nm [6], myosin VI can take 226nm/36nm ≈ 6.3 steps on average before releasing actin. Furthermore, because of the almost free diffusion along actin at near stall force, it is easy to understand why myosin VI exhibits remarkable oscillatory behavior with rapid transitions between two bound positions [6]. However, because there is always one head bound to actin, myosin V occasionally steps backward at high load [1].

In summary, we propose a new hand-over-hand model for the processive movement of myosin V and myosin VI that relies on chemical, mechanical, and electrical couplings. Their moving directions are determined by the relative orientations between their two heads at equilibrium and the heads' binding orientations on actin; while their mean step sizes are determined by the pseudo-repeat (~36 nm) of actin helix. Similar to kinesin, one head remains bound to actin during the moving period in one step for myosin V, which is like human walking; whereas for myosin VI, both heads are detached from actin during the moving period in one step, which is somewhat like human running. The processive movement does not need to rely on the two heads' coordination in their ATPase and mechanical cycles. Rather, the ATPase activities at the two heads are independent. The much higher ATPase rate at the trailing head compared with that at the leading head naturally makes myosin V walk processively, with one ATP generally being hydrolyzed per step. The actin-activated release of ADP makes myosin VI walk one step by consuming one ATP. This model is consistent with the measured pathway of myosin ATPase and the structural observations by electron microscopy. In particular, the model can give a very natural explanation of the similar dynamic behaviors but opposite moving directions of myosin VI and mutant myosin V. Using the model, the much shorter mean run length and the significantly broader distribution of step sizes of myosin VI (or mutant myosin V) compared with those of myosin V are well explained. The theoretical results of the dwell duration between steps versus load for different ATP concentrations and versus [ATP] at low load for myosin V are in agreement with previous experimental results.

This research is supported by the National Natural Science Foundation of China (Grant numbers: 60025516, 10334100).

**Appendix A**

Equation (1) can be rewritten in the following form

$$dx/dt = F_d + \tilde{f}(t) ,  \tag{A1}$$

where $F_d = (F_{driving} - F_{load})/\Gamma$. $\tilde{f}(t)$ satisfies $\langle \tilde{f}(t) \rangle = 0$, $\langle \tilde{f}(t)\tilde{f}(t') \rangle = 2D\delta(t-t')$, where $D = k_B T/\Gamma$. From Eq. (A1) we have the following Fokker-Planck equation

$$\frac{\partial W(x,t)}{\partial t} = -F_d \frac{\partial W(x,t)}{\partial x} + D \frac{\partial^2 W(x,t)}{\partial^2 x} . \tag{A2}$$

Using Eq. (5.2.157) in Ref. [36] and from Eq. (A2) we have

$$\psi(x) = \exp\left[\int_a^x (\frac{F_d}{D})dx\right] = \exp\left[\frac{F_d}{D}(x-a)\right], \tag{A3}$$

where the motion is supposed in the interval (*a*, *b*) and the barrier at *a* is reflecting and the barrier at *b* is absorbing. Using Eq. (5.2.160) in Ref. [36], the mean first-passage time can be written as

$$T(a) = 2\int_a^b \frac{dy}{\psi(y)} \int_a^y \frac{\psi(z)}{2D} dz. \tag{A4}$$

Integrating Eq. (A4) we obtain

$$T = \frac{1}{F_d}\left[L + \frac{D}{F_d}\left(\exp(-\frac{F_d}{D}L) - 1\right)\right], \tag{A5}$$

where for myosin V, $b - a \equiv L = 2d$, and for myosin VI, $L = d$.

**Appendix B**

As defined in the text, $g(t)dt$ represents the probability of the time interval between two successive collisions of ATP molecules with the ATP binding site of a MD at $t \sim t + dt$; $h(t)dt$ represents the probability of the ATP turnover time at $t \sim t + dt$, i.e., the time interval from the moment when an ATP molecule binds to the MD to the moment when $P_i$ and ADP are released. $g(t)$ and $h(t)$ are independent and they have the following forms

$$g(t) = \frac{a_1 a_2}{a_2 - a_1}\left[\exp(-a_1 t) - \exp(-a_2 t)\right], \quad (a_2 > a_1 > 0) \tag{B1}$$

$$h(t) = \frac{k_1 k_2}{k_2 - k_1}\left[\exp(-k_1 t) - \exp(-k_2 t)\right]. \quad (k_2 > k_1 > 0) \tag{B2}$$

From Eq. (B1) we can easily obtain the mean time $t_b$ of ATP binding is

$$t_b \equiv \frac{1}{k_b} = \frac{a_1 + a_2}{a_1 a_2}, \tag{B3}$$

where $k_b$ is the mean ATP binding rate. Similarly, from Eq. (B2) the mean ATP turnover time $t_c$ is

$$t_c \equiv \frac{1}{k_c} = \frac{k_1 + k_2}{k_1 k_2}, \tag{B4}$$

where $k_c$ is the mean ATP turnover rate.

In order to derive the mean ATPase time $\overline{T} \equiv 1/K$ (where $K$ is the mean ATPase rate), we first adopt $g(t) = \delta(t - \tau)$. In this case the mean ATPase time $t_0$ has the form

$$t_0 = \tau \int_0^\tau h(t)dt + 2\tau \int_\tau^{2\tau} h(t)dt + \cdots + n\tau \int_{(n-1)\tau}^{n\tau} h(t)dt. \quad (n \text{ is an integer and } n \to \infty). \tag{B5}$$

Substituting Eq. (B2) into Eq. (B5) we finally get

$$t_0 = \tau + \frac{k_2}{k_2 - k_1}\frac{\tau \exp(-k_1\tau)}{1 - \exp(-k_1\tau)} - \frac{k_1}{k_2 - k_1}\frac{\tau \exp(-k_2\tau)}{1 - \exp(-k_2\tau)}. \tag{B6}$$

Because $g(t)$ and $h(t)$ are independent, then we have the mean ATPase time $\overline{T}$

$$\overline{T} = \int_0^\infty t_0 g(t_0) dt_0. \tag{B7}$$

Substituting Eqs. (B1) and (B6) into Eq. (B7), we finally get

$$\overline{T} = \frac{a_1 + a_2}{a_1 a_2} + \frac{a_1 a_2}{a_2 - a_1}\frac{1}{k_2 - k_1}\left\{\frac{k_2}{k_1^2}\left(\zeta\left[2, \frac{k_1 + a_1}{k_1}\right] - \zeta\left[2, \frac{k_1 + a_2}{k_1}\right]\right) - \frac{k_1}{k_2^2}\left(\zeta\left[2, \frac{k_2 + a_1}{k_2}\right] - \zeta\left[2, \frac{k_2 + a_2}{k_2}\right]\right)\right\}, \tag{B8}$$

where $\zeta\left[2, \frac{k_1 + a_1}{k_1}\right]$, ......, $\zeta\left[2, \frac{k_2 + a_2}{k_2}\right]$ are zeta functions.

Since $1/t_b \equiv k_b = k_{b0}[\text{ATP}]$, from Eq. (B3) it is reasonable to assume

$$a_1 = a_1^{(0)}[\text{ATP}], \tag{B9}$$

$$a_2 = a_2^{(0)}[\text{ATP}]. \tag{B10}$$

Therefore,

$$k_{b0} = \frac{a_1^{(0)} a_2^{(0)}}{a_1^{(0)} + a_2^{(0)}}. \tag{B11}$$

For an approximation, the mean ATPase time $\bar{T}$ is usually taken in the following form in literatures

$$\bar{T} = t_b + t_c. \tag{B12}$$

From Eq. (B12) the well-known Michaelis-Menten equation is obtained

$$K \equiv \frac{1}{\bar{T}} = \frac{k_c[\text{ATP}]}{[\text{ATP}] + k_c/k_{b0}}. \tag{B13}$$

When there is a force exerted on the MD, the mean ATP turnover rate, $k_c$, and the mean ATP binding rate for unit [ATP], $k_{b0}$, follows the general Boltzman form [37]

$$k_c = \frac{k_c^{(0)}(1 + A_c)}{1 + A_c \exp(-F\delta/k_B T)}, \tag{B14}$$

$$k_{b0} = \frac{k_b^{(0)}(1 + A_b)}{1 + A_b \exp(-F\delta/k_B T)}. \tag{B15}$$

Similar to Eqs. (B9) and (B10), from Eq. (B4) and Eq. (B11), we have

$$k_1 = \frac{k_{10}(1 + A_c)}{1 + A_c \exp(-F\delta/k_B T)}, \tag{B16}$$

$$k_2 = \frac{k_{20}(1 + A_c)}{1 + A_c \exp(-F\delta/k_B T)}, \tag{B17}$$

$$a_1^{(0)} = \frac{a_{10}(1 + A_b)}{1 + A_b \exp(-F\delta/k_B T)}, \tag{B18}$$

$$a_2^{(0)} = \frac{a_{20}(1 + A_b)}{1 + A_b \exp(-F\delta/k_B T)}, \tag{B19}$$

where $\frac{k_{10} k_{20}}{k_{10} + k_{20}} = k_c^{(0)}$ and $\frac{a_{10} a_{20}}{a_{10} + a_{20}} = k_b^{(0)}$.

**Figures:**

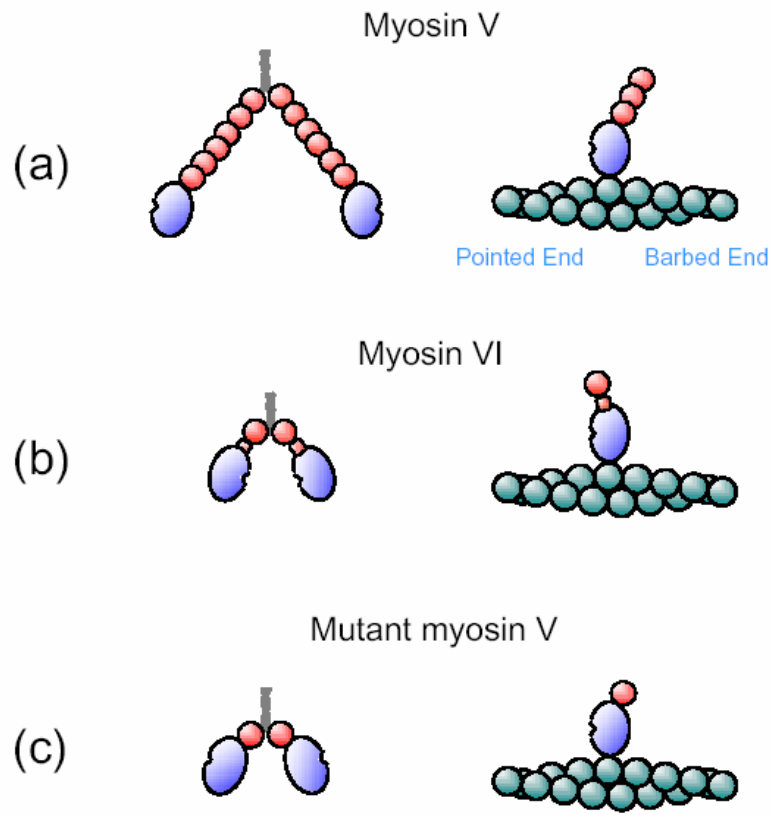

Fig. 1. Schematic drawing of the relative orientation between the two heads of the dimeric myosin at free state (left) and the binding orientation of its MD to actin filament (right). The MD's are in blue, neck domains in red, and coiled-coil in grey. The actin filament is in green. Although the actual myosin-binding site of actin is located between two actin monomers, here and in following figures binding is shown on only one monomer for simplicity. (a) Myosin V. Its neck domain has six IQ light chains bound with six calmodulin molecules (round). (b) Myosin VI. Its neck domain has only one IQ light chain bound one calmodulin molecule, and there is a unique insertion (square) between the neck and the MD. (c) Mutant myosin V with its neck domain truncated to only one-sixth of the native length [16].

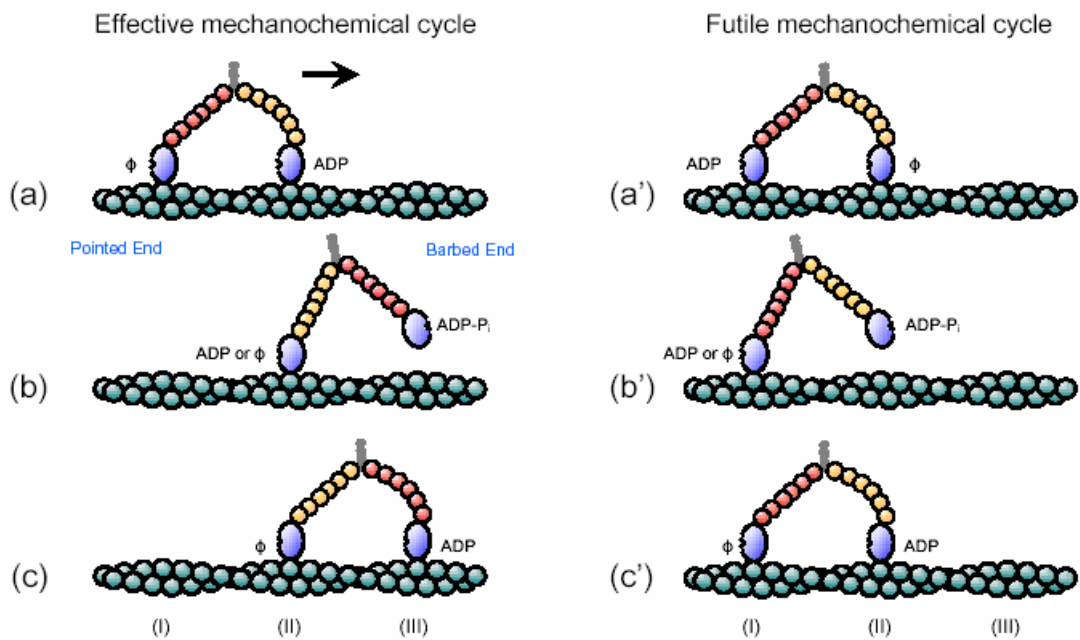

Fig. 2. Schematic illustrations of the proposed movement mechanism of myosin V. The moving direction is from pointed end to barbed end of actin. **Effective mechanochemical cycle:** (a) The cycle begins with both heads binding to actin (positions I and II). The trailing head (red neck) is in nucleotide-free state, the leading head (yellow neck) is in ADP–bound state. Note that the shapes of the two necks are determined by the internal elastic force and torque resulted from the orientation change of the leading MD from its equilibrium orientation as shown Fig. 1(a). (b) ATP binding or hydrolysis at the trailing head weakens the electrostatic binding force of actin due to the increase of the relative dielectric constant of the local solvent, and then driven by the internal elastic force and torque between the two heads the trailing head detaches from actin and subsequently moves to its equilibrium position. Now it is in ADP-$P_i$–bound state. (c) The detached head moves close to actin at position (III) via the electrostatic interactions. $P_i$ is rapidly released upon contacting of the detached head to actin and then the strong binding force drives it to its binding orientation. After ADP release in the new trailing head (yellow neck), one ATPase cycle is completed with one ATP being hydrolyzed per step. **Futile mechanochemical cycle:** (a') The cycle also begins with both heads binding to actin (positions I and II). The trailing head is in ADP–bound state and the leading head is in nucleotide-free state. (b') ATP binding or hydrolysis at the leading head promotes its detachment from actin at position (II) and subsequent movement towards the equilibrium position. Now it is in ADP-$P_i$–bound state. (c') The detached head rebinds to actin at position (II) after the original electrical property of local solvent is recovered. The leading head is in ADP–bound state while the trailing head is in nucleotide-free state. One ATP is hydrolyzed in this futile mechanical cycle, the end of which is the beginning of the effective mechanochemical cycle as shown in (a).

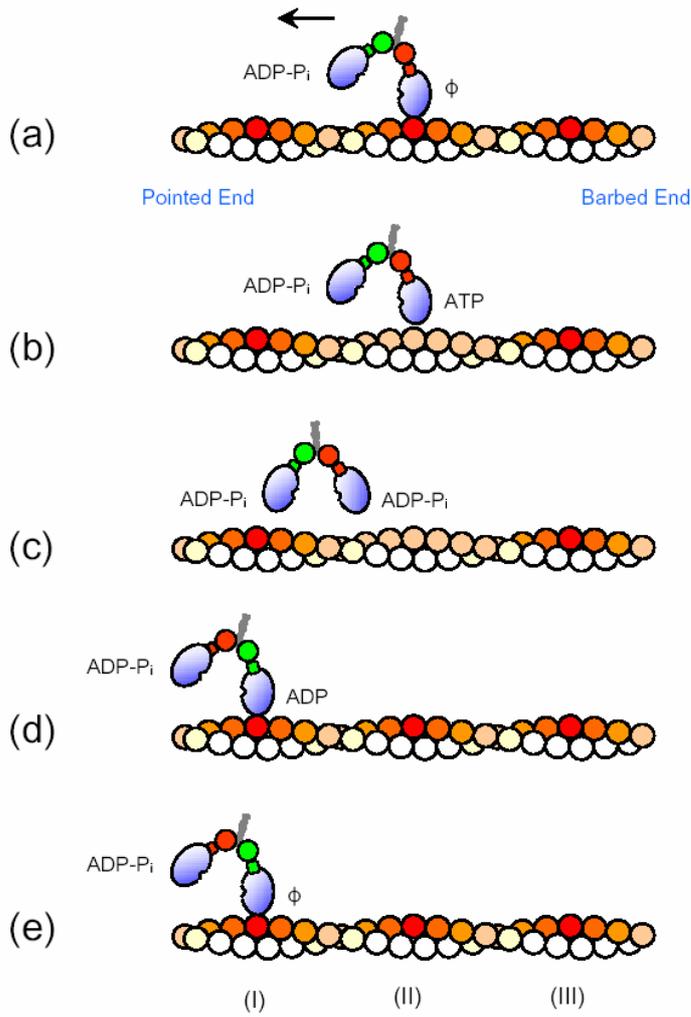

Fig. 3. Schematic illustrations of the proposed movement mechanism of myosin VI. The moving direction is from barbed end to pointed end. (a) The cycle begins with one nucleotide-free head (red neck) binding to actin at position (II). The detached head (green neck) is in ADP-$P_i$–bound state. The color spread of the actin monomers from red to shallow ones represents strong to weak variation of the electrostatic binding force of actin for a MD suspended above, as determined by the geometry of the actin filament [32, 33]. (b) ATP binding or hydrolysis at the bound head reduces the binding force of actin, and then driven by $F_E$ (see text) the bound head is detached from actin. Note that the binding force of the actin subunit from which the red-neck head has just detached and that of the nearby subunits become weak due to the increase of the relative dielectric constant of the local solvent. (c) Myosin VI rapidly moves toward the pointed end of actin driven by $F_E$. The green-neck head comes close to some strong binding site at position (I). (d) The green-neck head binds to actin with rapid release of $P_i$, the red-neck head going to its equilibrium position. After $\varepsilon_r^{(local)}$ of the local solvent near (II) relaxes to $\varepsilon_r$ of the solvent, the binding force of actin near (II) is recovered. (e) With ADP release, the mechanical cycle is completed. One ATP is hydrolyzed per step.

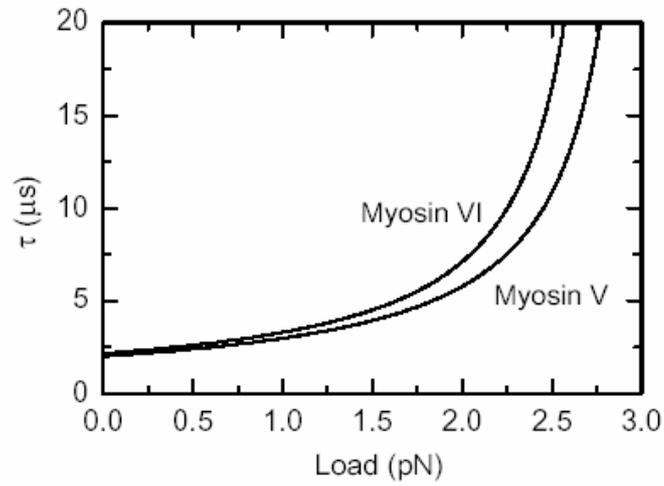

Fig. 4.  Moving time $\tau$ in one step versus load for myosin V and myosin VI.

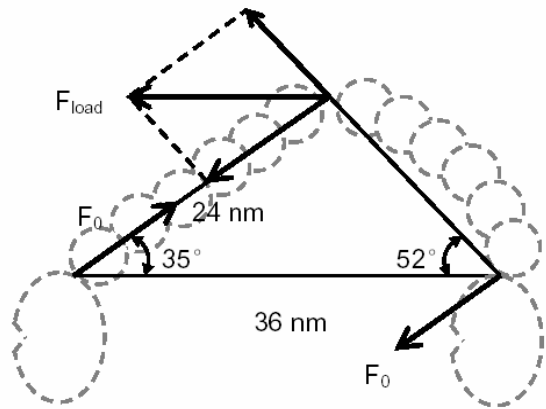

Fig. 5.  Geometry used to calculate forces acted on MD's of myosin V when there is a load $F_{load}$. $F_0$ is the internal elastic force, *i.e.*, the force acted on the MD's when there is no load.

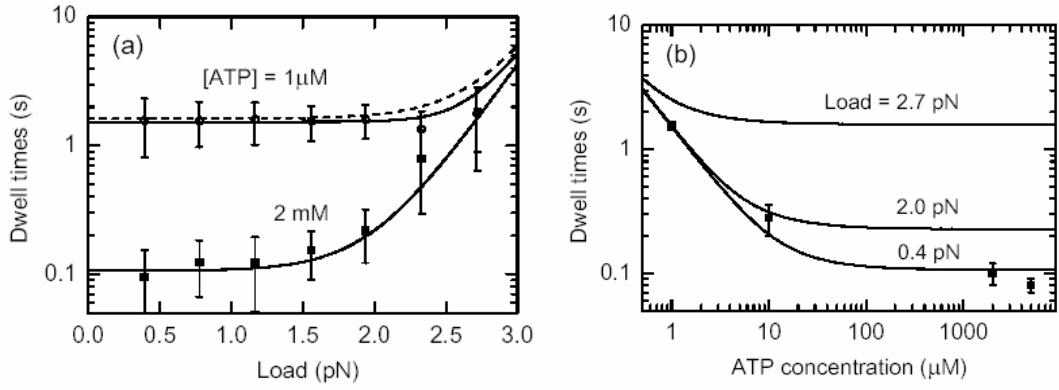

Fig. 6. Mean dwell duration $t_d$ of myosin V: (a) as a function of load at different ATP concentrations; (b) as a function of [ATP] under different loads. Experimental values (circles and squares) are redrawn from Mehta *et al.* [1]. Solid curves are calculated by using Eqs. (5)-(8a). Dashed curve is calculated by using Eqs. (3), (7), and (8a). The parameter values: $\delta = 18.3$ nm, $k_{10} = 0.25$ s$^{-1}$, $k_{20} = 3$ s$^{-1}$, $A_c = 40$, $a_{10} = 0.72$ μM$^{-1}$s$^{-1}$, $a_{20} = 3.6$ μM$^{-1}$s$^{-1}$, $A_b = 0.1$. The elastic force between the two heads is taken as $F_0 = 3$ pN.

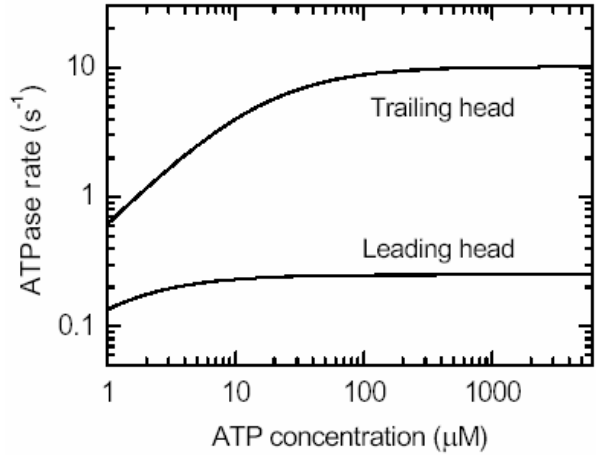

Fig. 7. ATPase rates at the trailing head and the leading head under no load for myosin V. The parameter values are the same as in Fig. 6.